\documentclass[pre,twocolumn,showpacs]{revtex4}
\usepackage{epsfig}
\usepackage{graphics}

\begin{document}

\title{Common Scaling Patterns in Intertrade Times of U. S. Stocks}

\author{Plamen~Ch.~Ivanov$^{1}$, Ainslie~Yuen$^{2}$, Boris~Podobnik$^{3}$, Youngki~Lee$^{4}$}

\affiliation{
$^{1}$Center for Polymer Studies and Department of Physics,
  Boston University, Boston, MA 02215\\
$^{2}$Signal Processing Laboratory, Department of Engineering, Cambridge University, UK\\
$^{3}$ Faculty of Civil Engineering, University of Rijeka, Rijeka, Croatia\\
$^{4}$ Yanbian University of Science and Technology, Yanji City, Jilin Province, China 133000}

\begin{abstract}
We analyze the sequence of time intervals between consecutive stock trades of thirty companies representing eight sectors of the U. S. economy over a period of four years. For all companies we find that: (i) the probability density function of intertrade times may be fit by a Weibull distribution; (ii) when appropriately rescaled the probability densities of all companies collapse onto a single curve implying a universal functional form; (iii) the intertrade times exhibit power-law correlated behavior within a trading day and a consistently greater degree of correlation over larger time scales, in agreement with the correlation behavior of the absolute price returns for the corresponding company, and (iv) the magnitude series of intertrade time increments is characterized by long-range power-law correlations suggesting the presence of nonlinear features in the trading dynamics, while the sign series is anti-correlated at small scales. Our results suggest that independent of industry sector, market capitalization and average level of trading activity, the series of intertrade times exhibit possibly universal scaling patterns, which may relate to a common mechanism underlying the trading dynamics of diverse companies. Further, our observation of long-range power-law correlations and a parallel with the crossover in the scaling of absolute price returns for each individual stock, support the hypothesis that the dynamics of transaction times may play a role in the process of price formation.
\end{abstract}
\date{\today}  

\maketitle

\begin{figure}

\centering
\includegraphics[width=8.5cm]{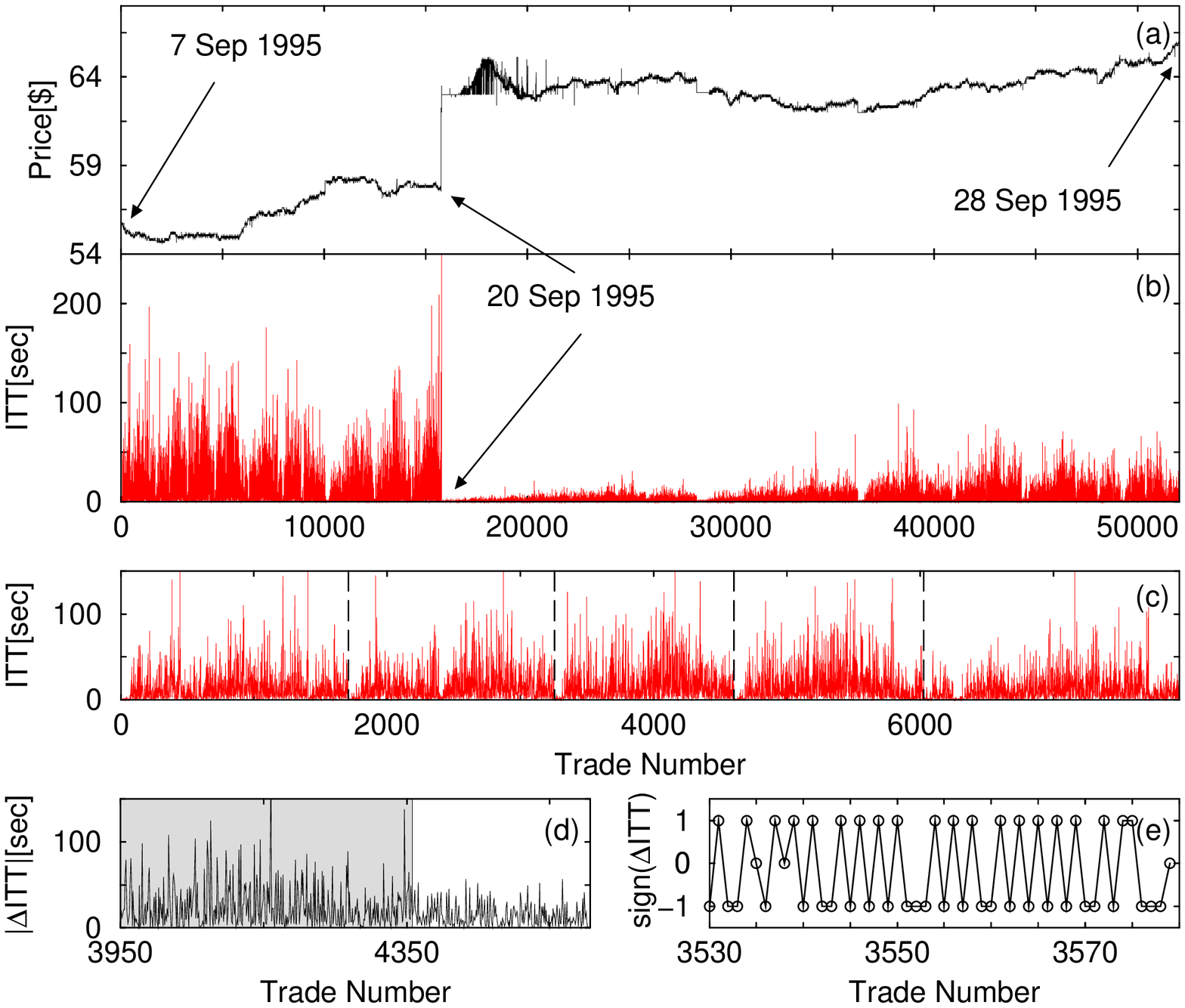}
\caption{ 
(a) Price of AT\&T stock over three weeks in September 1995 ($5.2 \times 10^{4}$ trades). On 20 Sept. 1995, AT\&T announced their intent to restructure into three separate companies, leading to a jump in the stock price. The price fluctuations exhibit a relaxation time of less than a day following this event. (b) Intertrade times (ITT) of AT\&T stock over the same period. Data exhibits highly heterogeneous structure with most of the trades concentrated in the third week. The relaxation time of the ITT response following the price jump is much longer than the relaxation time of the price fluctuations. (c) ITT data over the week beginning 7 Sept. 1995 --- trading days typically have short intertrade times at the open and close of business, with longer intertrade times in between. (d) Magnitude series of the increments $\Delta$ITT of consecutive intertrade times. Patches of more ``volatile" increments with large magnitude (shaded area) are followed by patches of less volatile increments with small magnitude, suggesting persistent behavior, in accordance with our finding (Fig.~\ref{fig.5}a). (e) Sign series of the increments of consecutive intertrade times. The apparent strong alternation between $+1$ and $-1$ is consistent with our finding of anti-persistent behavior at small scales (Fig.~\ref{fig.5}b).}
\label{fig.1}
\end{figure}

Investigations of price dynamics of financial assets and indices have long been the key focus of economic research \cite{Bachelier00,Mandelbrot63, Akgiray89,Dothan90,Koedijk90,Lux96,Campbell97,Vandewalle97,Gopi98,Vandewalle98,Lee98,Ivanova99,Plerou99,Mantegna00,Podobnik00,Bouchaud00, Muzy01,Ivanov01,Dacorogna01,Podobnik01,Kertesz02,Roehner02,Sornette03}. Recent studies, however, have turned to the information offered by other aspects of the trading process such as volume of shares traded at each transaction \cite{Jones94,Gopi00} or number of trades in a unit time \cite{Plerou00,Bonanno00}, and their possible relation to price formation \cite{Campbell97,Easley92,Engle98,Grammig02,Lillo03}. Empirical observations suggest a relationship between price and trading activity (Fig.~\ref{fig.1}). In addition, the impact of a significant price change on the level of trading activity can persist for much longer than the corresponding effect on the level of price fluctuations (Fig.~\ref{fig.1}a,b). These features suggest that information may be contained in the structure and temporal organization of trading activity, and that a close analysis of trading dynamics may offer quantitative insight into the complex mechanism driving price fluctuations \cite{Easley92,Bouchaud00}.

Recent studies have examined trading activity as measured by the average number of trades in a unit time \cite{Plerou00,Bonanno00}. However, aggregation into uniform time intervals may affect the analysis, since choosing a short unit time interval may result in many points with none or very few trades, artificially altering the heteroskedasticity of the process, while using a long unit time interval averages out multiple transactions, and the fine timing structure of the data can be lost \cite{Engle98}. To understand the dynamics of market activity on a trade-by-trade level, we consider the series of time intervals between consecutive trades, the intertrade times (ITT). Only few empirical studies of ITT have previously been carried out, examining a single actively traded stock over a period of a few months \cite{Engle98,Jasiak99,Golia01,Raberto02}, rarely traded nineteenth century stocks \cite{Sabatelli02}, or foreign exchange transactions \cite{Takayasu02,Marinelli01}.

\begin{table*}

\caption{ Descriptive statistics of the thirty U. S. stocks studied over the period 4 Jan.~93~-~31 Dec.~96. We include only intertrade times occurring 
during NYSE trading hours from 9.30am until 4pm EST, excluding public holidays 
and weekends. The period considered covers 1010 trading days. $\langle$M.C.$\rangle$ represents average market capitalization over the period in billions of U. S. dollars. $\langle$ITT$\rangle$ is the average intertrade interval over the period. $\alpha_{1}$ and $\alpha_{2}$ indicate the values of the scaling exponent characterizing power-law correlations in ITT and $|\Delta ITT|$  over small and large times scales. $\alpha_{1}$ is computed in the scaling range $1 \times 10^{-5}$ to $5 \times 10^{-4}$ for the first seventeen companies which have substantially less than $10^{6}$ trades, and from $4 \times 10^{-6}$ to $2 \times 10^{-4}$ for the remaining companies. $\alpha_{2}$ is computed in the scaling range $3 \times 10^{-3}$ to $10^{-1}$ for all companies (Fig.~\ref{fig.3}b,c). }

\label{Table 1}
\begin{centering}
\begin{ruledtabular}
\begin{tabular}{lccrccccc} 
Company&Symbol& $\langle$M.C.$\rangle$ &Number&$\langle$ITT$\rangle$&$\alpha_{1}$&$\alpha_{2}$&$\alpha_{1}$&$\alpha_{2}$\\ 
 &  & (\$$10^{9}$)& of trades& (sec)& ITT & ITT & $|\Delta$ITT$|$& $|\Delta$ITT$|$\\ \hline

Sprint Corp. & FON & 12.4 & 362851  & 64 & 0.64 & 0.95 & 0.72 & 0.94\\
Union Carbide Corp. & UK & 4.4 & 387273  & 60 & 0.65 & 0.96 & 0.72 & 0.95\\
Morgan JP \& Co. & JPM & 13.7 & 401213  & 58 & 0.61 &0.89 & 0.68 & 0.88\\
Dow Chemical Co. & DOW & 18.4 & 411258  & 57 & 0.62 &0.94 & 0.69 & 0.94\\
Chase Manhattan Corp. & CMB & 7.3 & 448801  & 52 & 0.66 & 0.94 & 0.71 & 0.94\\
3M & MMM & 24.8 & 449462  & 52 & 0.62 & 0.85 & 0.68 & 0.85\\
Texaco & TX & 18.4 & 457081  & 51 & 0.62 & 0.88 & 0.68 & 0.87\\
Archer Daniels Midland & ADM & 9.0 & 468148  & 50 & 0.63 & 0.98& 0.68 & 0.97\\
Eli Lilly \& Co. & LLY & 22.4 & 514899  & 45 & 0.65 & 0.94 & 0.68 & 0.94\\
Sara Lee Corp. & SLE & 13.4 & 527814  & 44 & 0.62 & 0.94 & 0.66 & 0.93\\
Du Pont & DD & 39.0 & 543724  & 43& 0.62 &0.88 & 0.66 & 0.88\\
Fed. Natl. Mort. Assoc. & FNM & 26.5 & 627313  & 37 & 0.64 & 0.89 & 0.67 & 0.88\\
Citicorp & CCI & 22.6 & 677484  & 34 & 0.66 & 0.92 & 0.69 & 0.92\\
Pfizer & PFE & 29.7 & 689705  & 34 & 0.64 & 0.89 & 0.67 & 0.89\\
Abbott Laboratories & ABT & 28.2 & 691877  & 34 & 0.64 & 0.88& 0.67 & 0.87\\
Boeing & BA & 19.9 & 728779 & 32& 0.65 &0.94& 0.67 & 0.94\\
Exxon & XON & 87.5 & 750298 & 31 & 0.63 &0.88 &0.65 & 0.86\\
Johnson \& Johnson & JNJ & 41.6 & 1001549 & 23 & 0.63 & 0.92 & 0.71 & 0.92\\
Home Depot & HD & 20.7 & 1103037 & 21 & 0.62 & 1.03 &0.68 & 1.04\\
Bristol Myers Squibb & BMY & 35.1 &1121714 & 21 & 0.62 & 0.91 & 0.68 & 0.90\\
General Motors Corp. & GM & 35.6 & 1130452 & 21 & 0.64 & 0.95 & 0.69 & 0.95\\
Chrysler Corp. & C & 18.4 & 1231979 & 19 & 0.65 & 0.95 & 0.70 & 0.95\\
Coca Cola & KO & 77.0 & 1244660 & 19 & 0.63 & 0.99 &0.68 & 0.98\\
General Electric & GE & 101.7 & 1374682 & 17 & 0.61 &0.90 & 0.66 & 0.91\\
Philip Morris & MO & 60.2 & 1527659 & 15 & 0.64 & 1.06 & 0.66 & 1.06\\
IBM & IBM &  45.4 & 1677319 & 14 & 0.65 &0.94 & 0.67 & 0.94\\
AT \&T & T & 82.1 & 1689767 & 14  & 0.64 &1.04 & 0.66 & 1.04\\
Wal Mart & WMT & 58.2 & 1794160 & 13 & 0.66 &1.01 & 0.68 & 1.01\\
Merck \& Co. & MRK & 56.9 & 2055443 & 11 & 0.65 &0.94 & 0.66 & 0.94\\ 
Motorola & MOT & 30.4 & 2204059 & 11 & 0.65 & 1.04& 0.66 & 1.04\\ 

\end{tabular}
\end{ruledtabular}
\end{centering}
\end{table*}

Here we empirically investigate the statistical and scaling properties of ITT over extended periods of time. In particular, we hypothesize that trading dynamics may carry features independent of individual company characteristics such as industry sector, level of trading activity and market capitalization. We examine thirty stocks listed on the New York Stock Exchange (NYSE) from eight sectors of the US economy: Technology/Communications(4), Pharmaceutical(6), Retail\&Food(8), Automotive(2), Oil(2), Aerospace(1), Financial(4) and Chemicals(3).  We study the time intervals between consecutive stock trades over a period of four years --- Jan.~93 till Dec.~96 --- as recorded in the Trades and Quotes (TAQ) database (NYSE, New York, 1993). The thirty companies vary in their average market capitalization and exhibit different levels of trading activity with different numbers of trades over this period (Table~\ref{Table 1}).

\begin{figure}
\centering

\includegraphics[width=7.6cm]{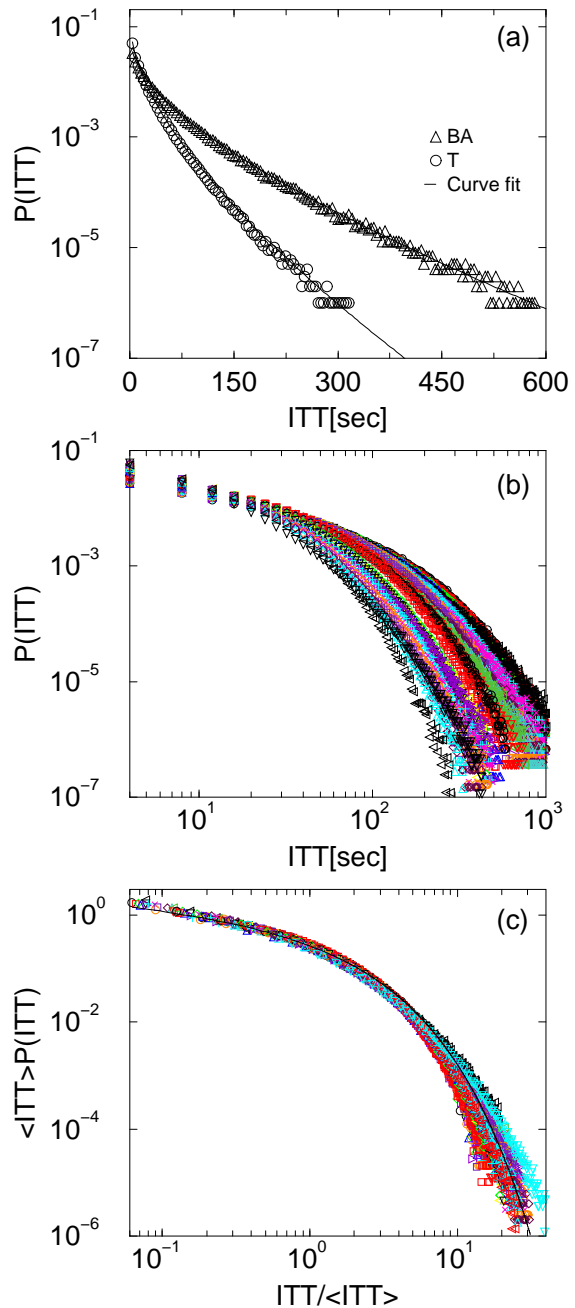}

\caption{(a) Probability density functions with Weibull fits (solid lines) of intertrade times (ITT) over the period Jan.~93~-~Dec.~96 for two company stocks: Boeing (BA) fit with parameters $\delta=0.73,\tau=27$, and AT\&T (T) fit with $\delta=0.70,\tau=11$ (Eq.~\ref{Eqn 1}). We use 4 sec bins, where ITT values in [2,6) are centered at 4 sec, values in [6,10) are centered at 8 sec etc. (b) Probability density functions of ITT for thirty U.S. stocks over the same period as in (a) with increasing number of trades from top to bottom (Table~\ref{Table 1}). (c) Same probability distributions as in (b) after rescaling $P($ITT$)$ by $\langle$ITT$\rangle$ and ITT by 1/$\langle$ITT$\rangle$. This rescaling is equivalent to that described in Eq.~\ref{Eqn 3} as $P($ITT$)\equiv P(x,\tau)$ and $\tau \sim$$\langle$ITT$\rangle$. Data points collapse onto a single scaled curve. The solid line represents a Weibull fit to the data points with $\delta=0.72,\tau=0.94$. }
\label{fig.2}
\end{figure}

We first study the probability density function of ITT. The distribution changes as companies with more frequently traded shares have a higher peak at shorter intertrade times, while more rarely traded companies have tails extended over larger intertrade times (Fig.~\ref{fig.2}).

We find that the individual probability distributions for all thirty companies are well fit by a generalized homogeneous form --- the Weibull distribution
\begin{equation}
P(x,\tau) = \frac{\delta}{\tau}\left(\frac{x}{\tau}\right)^{\delta-1}exp\left[-\left(\frac{x}{\tau}\right)^{\delta}\right],
\label{Eqn 1}
\end{equation} 
where $\delta$ is the stretched exponent (or shape parameter) and $\tau$ is the characteristic time scale (Fig.~\ref{fig.2}a) \cite{Reiss01}. Studies of short ITT sequences --- IBM stock trades during Nov.~90 to Jan.~91 \cite{Engle98} and GE stock trades during Oct.~99 \cite{Raberto02} --- suggest stretched exponential behavior for the tails of the probability distributions, in agreement with Eq.~\ref{Eqn 1} \cite{footnote_bonds}. The power-law prefactor in the Weibull form accounts for the steeper (relative to the stretched exponential) trend in the distribution at small ITT values. This functional form is markedly different to the power-law form of the distribution of number of stock trades in a unit time reported previously \cite{Plerou00}. 

Since different companies have different average intertrade intervals $\langle$ITT$\rangle$ (Table~\ref{Table 1}), they are also characterized by a different parameter $\tau$. A function $P(x,\tau)$ is a generalized homogeneous function if there exist two numbers $r$ and $s$, termed scaling parameters, such that, for all positive values of the parameter $\lambda$,
\begin{equation}
P(\lambda^{r}x,\lambda^{s}\tau)=\lambda P(x,\tau).
\label{Eqn 2}
\end{equation}
Generalized homogenous functions are defined as solutions of this functional equation. $P(x,\tau)$ satisfies Eq.~\ref{Eqn 2} with $r=-1$ and $s=-1$. Data collapsing is an important property of generalized homogeneous functions: instead of data for $P(x,\tau)$ falling on a family of curves, one for each value of $\tau$, data points collapse onto a single curve given by the scaling function 
\begin{equation}
\widetilde{P}(\widetilde{x}) \equiv \widetilde{P}(\frac{x}{\tau},1)=\tau P(x,\tau),
\label{Eqn 3}
\end{equation}
where the number of independent variables is reduced by defining the scaled variable $\widetilde{x} \equiv x/\tau$.

To test the hypothesis that there is a possibly universal structure to the intertrade time dynamics of diverse companies, we rescale the distributions. We find that for all companies, data conform to a single scaled plot --- ``data collapse" (Fig.~\ref{fig.2}c) \cite{Stauffer96}. Such behavior is a hallmark of scaling, and is typical of a wide class of physical systems with universal scaling properties \cite{Bunde94}.

\begin{figure*}
\centering

\includegraphics[width=17.5cm]{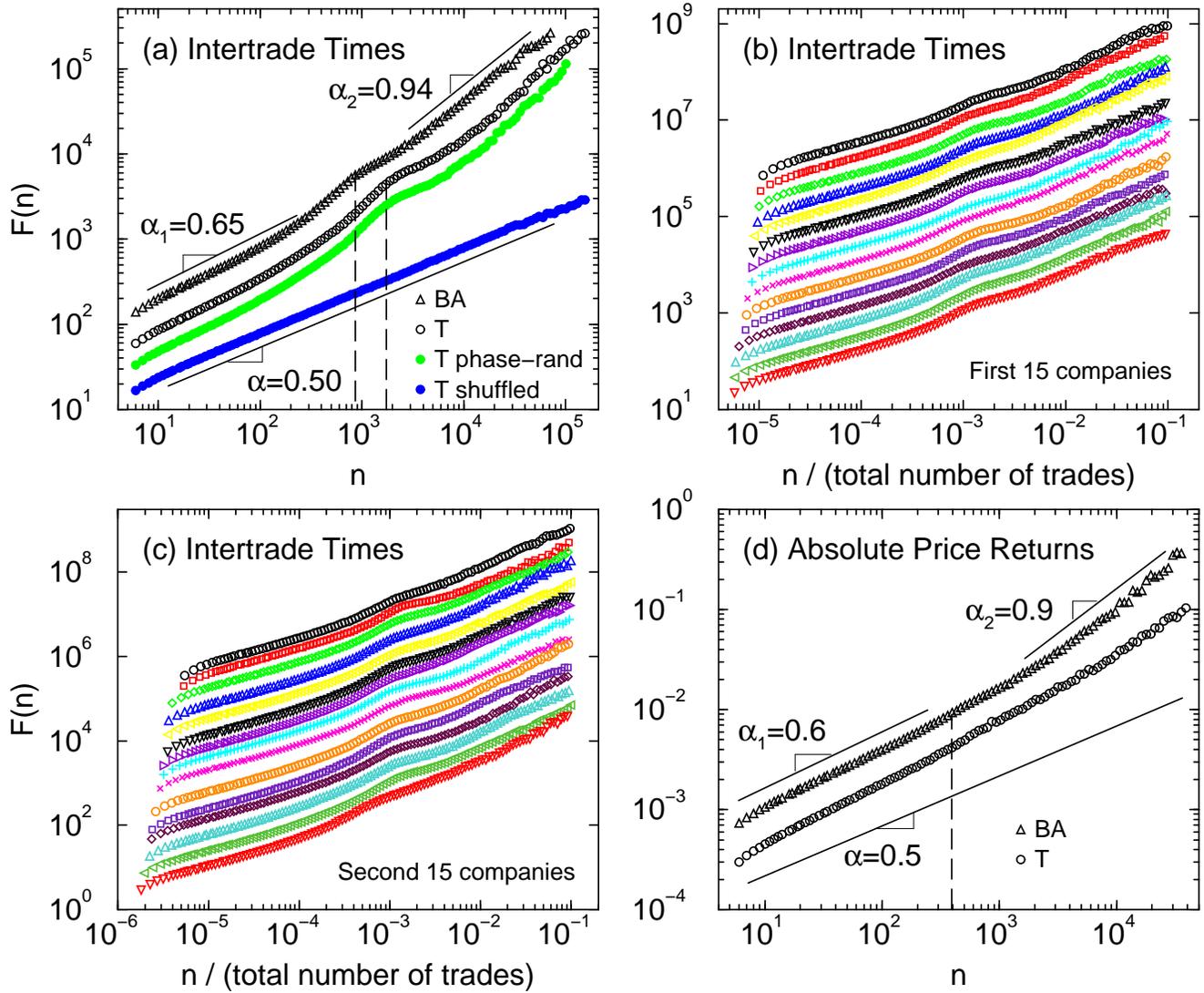}

\caption
{
(a) Root mean square fluctuation, $F(n)$, for intertrade times (ITT) for companies Boeing (BA) and AT\&T (T) obtained using DFA-2 analysis. Here, $n$ indicates the time scale in trade number. Both series exhibit long-range power-law correlations with a pronounced crossover to larger exponent at scales above one trading day. The average daily number of trades for each company is marked by a dashed line. As expected, the scaling properties of the ITT series remain unchanged after the Fourier-phase randomization, while the shuffled ITT series is characterized by exponent $\alpha =0.5$ as for uncorrelated (white) noise. Curves are vertically offset for clarity. $F(n)$ for the ITT series of (b) the first group of 15 companies and (c) the second 15 companies as ordered in Table~\ref{Table 1}. Curves are vertically shifted with approximately equal spacing and the crossovers are aligned by rescaling the time scale $n$ by the total number of trades for each company in the period 4 Jan.~93~-~31 Dec.~96. All companies show a remarkably common scaling behavior. (d) $F(n)$ for the time series of absolute logarithmic price returns computed per minute for Boeing and AT\&T. Here, $n$ indicates the time scale in minutes. The vertical dashed line marks a crossover at $\approx$ 390 minutes --- a typical trading day. Both companies exhibit scaling behavior similar to that observed for their respective ITT series, with a greater degree of correlation over scales above one trading day. }
\label{fig.3}
\end{figure*}

Next, we investigate the temporal organization of ITT. Empirical observations show that like many other financial time series, the ITT data exhibit nonstationary behavior and complex variability with a superposed pattern of daily activity (Fig.~\ref{fig.1}). Patches of inactive trading are often followed by patches of more active trading within a trading day. Similar patterns can be observed on a daily, weekly and even monthly basis, independent of the average level of trading activity for a company. Such observations suggest that there may be a self-similar, fractal organization in the sequence of ITT over a broad range of time scales. To test this hypothesis we apply the detrended fluctuation analysis (DFA) method \cite{Buldyrev93,Peng94}. The DFA method can accurately quantify fractal features in ITT, as it permits the detection of long-range correlations embedded in nonstationary time series, and avoids the spurious detection of apparent long-range correlations that are an artifact of trends in the data \cite{Hu01}.

The DFA method consists of the following steps. We first integrate the ITT series to construct the profile
$Y(k)=\sum^{k}_{i=1}(ITT_{i}-\left\langle ITT\right\rangle)$
where $\left\langle ITT \right\rangle$ is the series mean. Next,
we partition the profile $Y(k)$ into non-overlapping segments of length $n$ (number of consecutive intertrade intervals) and fit the local trend in each segment with a least-squares polynomial fit. We then detrend the profile $Y(k)$ by subtracting the local polynomial trend in each segment of length $n$, and we calculate the root mean square fluctuation $F(n)$ for the detrended profile. For order-$l$ DFA (DFA-$1$ if $l=1$, DFA-$2$ if $l=2$, etc.) a polynomial function of order $l$ is applied for the fitting of the local trend in each segment of the profile $Y(k)$. This procedure is repeated for different scales $n$. A power-law relation $F(n)\sim n^{\alpha}$ indicates the presence of scaling in the ITT series. Thus the fluctuations in the ITT can be characterized by scaling exponent $\alpha$, a self-similarity parameter that quantifies the fractal power-law correlation properties of the signal. The scaling or correlation exponent $\alpha$ is related to the autocorrelation function exponent 
$\gamma$ ($C(n)\sim n^{-\gamma }$ when $0<\gamma<1$)
and to the power spectrum exponent $\beta$ ($S(f)\sim 1/f^\beta$)
by $\alpha=1-\gamma/2=(\beta+1)/2$~\cite{Buldyrev93,Ben-Avraham00}.
A value of $\alpha =0.5$ indicates that
there are no correlations and the signal is uncorrelated (white noise).
If $\alpha <0.5$ the signal is said to be {\it  anti-correlated}, meaning that 
large values are more likely to be followed by small values. If 
$\alpha >0.5$ the signal is correlated and exhibits persistent behavior, 
meaning that large values are more likely to be followed by large 
values and small values by small values. The higher the value of $\alpha$, the stronger the correlations in the signal.

Before performing the DFA analysis we pre-process the data by excluding all outliers in the ITT series exceeding ten times the standard deviation above zero. This naturally excludes large ITT values caused by unusual closures inside a trading day, as well as data entry errors. This procedure results in the removal of less than $0.06\%$ of all data points. In addition, the split transactions that arise when the volume of an order must be matched by several opposing orders often results in a number of transactions having an execution time separated by less than a second. Given the one-second resolution of the recordings, these transactions result in between $4-16\%$ zero intertrade intervals for individual company datasets, with a mean value of $7.3\%$ for the whole database. Removal of all outliers and all zero intertrade times does not significantly affect the results of the DFA analysis for positively correlated signals \cite{Chen02}. 

We find that the ITT series for all companies exhibit long-range power-law correlations over a broad range of time scales from several trades to hundreds of thousands of trades characterized by a correlation exponent $\alpha >0.5$ (Fig.~\ref{fig.3}, Table~\ref{Table 1}). This is consistent with the empirical observation that long segments of high trading activity (small values of ITT) may follow long segments of less active trading (large values of ITT) (Fig.~\ref{fig.1}b). We find that this scaling behavior is independent of the specific company and its market capitalization, the average level of trading activity and the industry sector. To confirm the presence of such strong persistent behavior we shuffle the data and obtain white noise behavior with exponent $\alpha=0.5$, significantly different from the behavior of the original ITT series (Fig.~\ref{fig.3}a).

For all companies we observe two scaling regimes, one at short time scales ranging from several trades to a few thousand trades within a trading day, followed by a second regime ranging from thousands to hundreds of thousands of trades corresponding to time scales from days to almost a year (Fig.~\ref{fig.3}a-c). These two scaling regimes are separated by a bump in the scaling curve $F(n)$ due to the periodic daily pattern in trading activity (Fig.~\ref{fig.1} and Fig.~\ref{fig.3}). Such periodic trends superposed on power-law correlated signals do not affect the value of the DFA scaling exponent \cite{Hu01}.

\begin{figure}
\centering

\includegraphics[width=8cm]{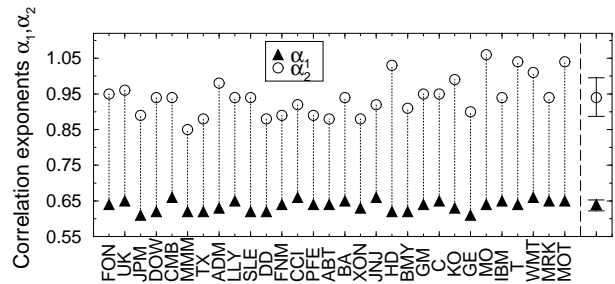}
\caption
{ Values of the DFA correlation exponents for a diverse group of companies (Table~\ref{Table 1}). The ITT series of all companies exhibit systematically weaker correlations over time scales less than a trading day (small values for the scaling exponent $\alpha_{1}$), and stronger correlations over time scales above one trading day (larger values for $\alpha_{2}$). Group averages and standard deviations are shown to the right of the panel. }
\label{fig.4}
\end{figure}

Since different companies exhibit daily patterns in the ITT series characterized by a different number of trades per day, we align the scaling regimes for all companies by normalizing the scale $n$ by the total number of trades in each series (Fig.~\ref{fig.3}b,c). Remarkably we find that all companies have common scaling behavior characterized by a correlation exponent $\alpha_{1}=0.64 \pm 0.02$ (group mean $\pm$ std. dev.) at time scales within a trading day, and by correlation exponent $\alpha_{2}=0.94 \pm 0.05$ at time scales larger than a trading day (Fig.~\ref{fig.4}). Furthermore, we find that a higher value of $\alpha_{1}$ for a given company is usually accompanied by a higher value of $\alpha_{2}$, resulting in a systematic difference between the scaling exponents of $\alpha_{2} - \alpha_{1}=0.30 \pm 0.05$ (group mean $\pm$ std. dev.). 

\begin{figure}
\centering
\includegraphics[width=7cm]{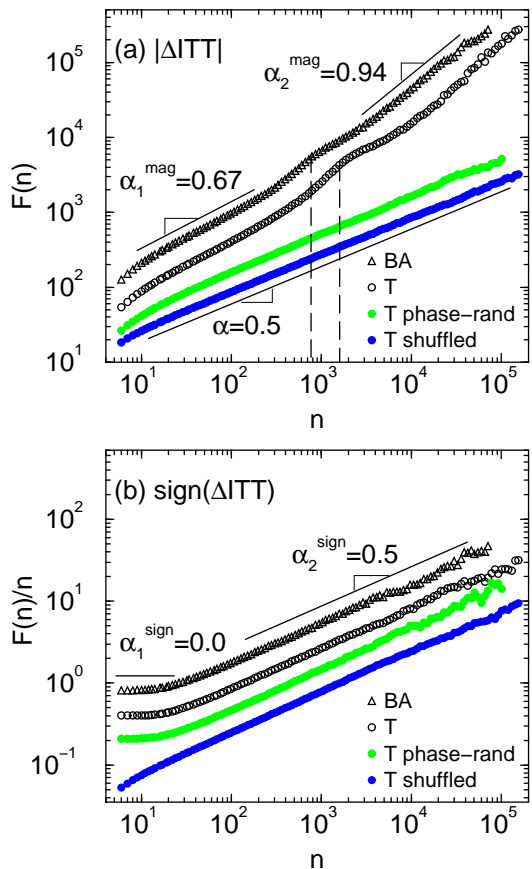}

\caption
{(a) Root mean square fluctuation, $F(n)$, for the magnitude series of the increments $\Delta$ITT for companies Boeing (BA) and AT\&T (T) obtained using DFA-2 analysis. Both series exhibit similar scaling behavior to that found for the ITT series (Fig.~\ref{fig.3} a), with a crossover from lower to higher correlation exponent centered at the average daily number of trades for each company (vertical dashed line). In contrast, the magnitude series of the surrogate signal obtained by Fourier-phase randomization of the ITT series is uncorrelated with exponent $\alpha =0.5$ as observed for the shuffled ITT series. This change in the scaling (after Fourier-phase randomization) suggests that the magnitude series carries information about the nonlinear properties of the ITT \cite{Ashkenazy03}. Curves are vertically shifted for clarity. (b) Scaling of the sign series of $\Delta$ITT. Strongly anti-correlated behavior at short time scales is followed by uncorrelated behavior over larger time scales. This scaling behavior remains unchanged after Fourier-phase randomization of the ITT, suggesting that the sign series relates to the linear properties of the ITT series. We take sign($\Delta$ITT=0)=0, and we integrate the sign series before DFA analysis to accurately quantify strong anti-correlations \cite{Hu01,Chen02}. To account for this integration we measure the slope of $F(n)/n$.}
\label{fig.5}
\end{figure}

This correlation behavior of the ITT series is also surprisingly reminiscent of the scaling features of the absolute price return series \cite{Liu99}. For each company both series show (i) two scaling regimes separated by a crossover at time scales corresponding to one trading day, (ii) positive correlations within a trading day ($\alpha_{1}^{price}=0.59 \pm 0.01$, group mean $\pm$ std. dev.), and even stronger correlations ($\alpha_{2}^{price}=0.77 \pm 0.06$), over larger time scales, and (iii) very similar values of the DFA correlation exponents in the respective scaling regimes (Fig.~\ref{fig.3}a,d). Such parallels in the scaling of ITT and absolute returns at both short and long time scales suggest an intrinsic relation between trading activity and stock price formation on an individual company basis.

To better understand the temporal organization of trading dynamics and the nature of the observed power-law scaling, we decompose the ITT series into a magnitude and sign series of the increments $\Delta$ITT in the consecutive intertrade intervals. Since underlying market interactions determine the magnitude ($|\Delta$ITT$|$) and direction (sign($\Delta$ITT)) of the ITT fluctuations, we separately analyze the correlations in the magnitude and sign series. Previous work has demonstrated that signals with identical long-range power-law correlations can exhibit different time ordering for the magnitude and sign series \cite{Ashkenazy01}.

We find that for all companies the magnitude series exhibits long-range persistent behavior (Fig.~\ref{fig.5}a) with practically identical correlation exponents to the original ITT series --- $\alpha_{1}^{mag}=0.68 \pm 0.02$ (group mean $\pm$ std. dev.) over short time scales within a trading day and $\alpha_{2}^{mag}=0.94 \pm 0.06$ over large time scales. Correlation in the magnitude series indicates that an increment with large magnitude is more likely to be followed by an increment with large magnitude (Fig.~\ref{fig.1}d). In contrast we find that the sign series for all companies is strongly anti-correlated over short time scales with $\alpha_{1}^{sign}=0.04 \pm 0.02$ (group mean $\pm$ std. dev.) and is uncorrelated over large time scales with $\alpha_{2}^{sign}=0.50 \pm 0.01$ (Fig.~\ref{fig.5}b). Thus our results suggest an empirical ``rule" for the temporal organization of ITT fluctuations: a large positive increment in intertrade interval is followed by a large negative increment, and this holds over a broad range of time scales.

We next demonstrate that the scaling features of the magnitude and sign series are independent of those of the ITT series. We perform a Fourier transform on the ITT series, and preserve the Fourier amplitudes but randomize the Fourier phases. Then we take the inverse Fourier transform to create a surrogate signal. This procedure eliminates nonlinearities, preserving only the linear features (i.e., power spectrum) of the original ITT series \cite{Theiler92}. The surrogate (linearized) signal has the {\it same} two-point correlations as the original ITT series with practically identical correlation exponents $\alpha_{1}$ and $\alpha_{2}$ indicating long-range correlations (Fig.~\ref{fig.3}a). We find that the sign series derived from the surrogate signal shows scaling behavior virtually identical to that of the sign series derived from the original ITT series (Fig.~\ref{fig.5}b). However, the magnitude series derived from the surrogate (linearized) signal exhibits {\it uncorrelated} behavior --- a significant change from the strongly correlated behavior we find for the original magnitude series (Fig.~\ref{fig.5}a). Thus the increments in the surrogate signal exhibit different time ordering for the magnitude, and do not follow the empirical rule observed for the increments of the ITT series, although the surrogate signal follows a scaling law identical to the original ITT series. Further, our results suggest that the ITT series has nonlinear properties encoded in the Fourier phases and represented by the long-range correlations in the magnitude series. In contrast, the sign series relates to the linear properties of ITT.

In summary, we present an empirical study of intertrade time dynamics for a diverse group of stocks listed on the NYSE. Our findings suggest that a single, possibly universal, functional form defines the probability density of the intertrade times of each company. Further, we find a common scaling behavior in the temporal organization of trading, characterized by long-range power-law correlations within a trading day and by a crossover to even stronger correlations over scales of days, months and years. These scaling patterns appear independent of level of trading activity, market capitalization or industry sector, and thus may be inherent to the trading process. The two scaling regimes in the ITT and $|\Delta$ITT$|$ series may be a consequence of the time scales over which news is absorbed \cite{Easley92}. Since trading activity is influenced by information, there will be less coherence between intra-day trading as information takes time to disseminate, thus leading to a lower value of the correlation exponent $\alpha$. Over time scales greater than a day more information is available to investors, resulting in a transition to more coherent and thus more persistent behavior with a higher value of $\alpha$.
The universality of this behavior and our observation of a parallel with the crossover in the scaling of price fluctuations for each individual stock support the hypothesis that the dynamics of transaction times may play a role in the process of price formation, and may have implications for financial modeling based on continuous time random walks \cite{Montroll65,Scalas00,Mainardi00,Masoliver03},
stochastic subordinated-processes \cite{Clark73,Dacorogna93,Ghysels95,Mandelbrot97,Marinelli01} and
agent-based modeling \cite{Lux99,Cont00,Zawadowsk02,Daniels03} of market behavior.

\begin{acknowledgments}
We thank K. Hu, T. Lim and S. White for stimulating discussions. A. Y. thanks the Department of Engineering, Cambridge University and King's College, Cambridge for financial support.
\end{acknowledgments}


\begin{thebibliography}{0}
\bibitem{footnote_bonds}
Our observation of a Weibull distribution for intertrade times of U.S. stocks appears to be quite different from the distribution of intertrade times of European Government bonds where a Mittag-Leffler functional form with an extended power-law tail was reported \cite{Mainardi00}. While waiting times of up to $5 \times 10^{4}$ sec are comparatively frequent for European bonds, thus leading to extended tails, the longest intertrade times observed in our group of U.S. stocks are in the range of $10^{3}$ sec and are rare (Fig.~\ref{fig.2}b), leading to stretched exponential tails.
\end{thebibliography}
\end{document}